\documentclass[aps,pre,twocolumn,showpacs,showkeys,groupedaddress]{revtex4-2}

\usepackage{graphicx}
\usepackage{rotating}
\usepackage{amssymb}
\usepackage{amsmath}
\usepackage{epsfig}
\usepackage[normalem]{ulem}
\usepackage{bm}
\usepackage{color}

\date{\today}

\begin{document}

\title{Local control for the collective dynamics of self-propelled particles}

\author{Everton S. Medeiros}
\author{Ulrike Feudel}
\affiliation{Institute for Chemistry and Biology of the Marine Environment, Carl von Ossietzky University Oldenburg, 26111 Oldenburg, Germany}

\begin{abstract}
Utilizing a  paradigmatic model for the motion of interacting self-propelled particles, we demonstrate that local accelerations at the level of individual particles can drive transitions between different collective dynamics, leading to a control process. We find that the ability to trigger such transitions is hierarchically distributed among the particles and can form distinctive spatial patterns within the collective. Chaotic dynamics occur during the transitions, which can be attributed to fractal basin boundaries mediating the control process. The particle hierarchies described in this study offer decentralized capabilities for controlling artificial swarms.
\end{abstract}
\maketitle

\section{Introduction}

In nature, groups of self-propelled elements interact and spontaneously form collectives, or swarms, such as flocks of birds \cite{Ballerini2008, Bajec2009, Pearce2014}, schools of fish \cite{Partridge1982}, and colonies of bacteria \cite{Sokolov2007, Kearns2010}. The functioning of such formations requires a decentralized decision-making process that often relies on pertinent information, e.g., food location, available to only a few constituents \cite{Reebs2000}. Therefore, a collective decision is based on efficiently transferring information from local to global scales within these formations \cite{Couzin2005,Attanasi2014,Attanasi2015,Ziepke2022}. The success of such natural mechanisms motivates research toward enabling artificial systems to exhibit collective capabilities such as self-organization in space, navigation behaviors, and decision-making (see \cite{Brambilla2013,Schranz2020} and refs. therein). To accomplish these tasks, different strategies have been developed \cite{Soysal2005,Turgut2008,Ferrante2012,Francesca2014,Cavagna2015,Vasarhelyi2018,Firat2020,Edwards2020}. However, controlling the switching among different collective patterns via local interventions is an open problem in such applications. A step toward such accomplishment is understanding the boundaries that separate coexisting collective motion.

Here, we investigate this issue via computer simulations of a system of interacting self-propelled particles, a suitable model for swarms of artificial elements. As coexistent collective motion, we consider two types of behavior exhibited by the particles: a translational and a rotational state. We demonstrate that local accelerations --at the level of individual particles-- can trigger transitions between these two collective states. By applying large sets of different accelerations to each particle of the swarm in a given state, we observe that the accelerations that trigger transitions form subsets of various sizes, yielding an internal hierarchy among the particles. Remarkably, in low-density swarms, particles in the upper hierarchy form patterns of high control effectivity that remain inalterable for different swarm sizes and initial conditions. For denser swarms, we show that local controllability is still achievable by adjusting the energy balance of the particles. Although patterns of high control effectivity may be less readily distinguishable, the particle hierarchies persist. Moreover, we observe chaotic motion during the driven transitions, signaling the presence of unstable chaotic sets and associated fractal basin boundaries separating the states. These basin boundaries often extend over a significant portion of the system's state space, increasing the likelihood of inducing transitions between states. Consequently, fractal basin boundaries play a pivotal role in the success of our control approach. Finally, we examine a simplified model describing the interacting self-propelled particles to illustrate that homoclinic intersections can generate such fractal basin boundaries..

\section{Local control for collective states}

The equations describing the system with $N$ interacting self-propelled particles (SPPs) are given by:

\begin{eqnarray}
 \label{Eq_1_model}
  \frac{\partial \vec{r}_i}{\partial t} &=& \vec{v}_i,\\
 m\frac{\partial \vec{v}_i}{\partial t} &=& (\alpha - \beta|\vec{v}_i|^2)\vec{v}_i-\vec{\nabla}_iU(\vec{r}_i),
 \label{Eq_2_model}
\end{eqnarray}
where the vectors $\vec{r}_i$ and $\vec{v}_i$ are, respectively, the position and velocity of the $i$th particle in a plane with $i=1,\dots,N$. The constants $\alpha$ and $\beta$ account for velocity-dependent admission and dissipation of energy of the SPPs system, respectively. The pairwise interaction among the particles is given by the generalized Morse potential:
\begin{eqnarray}
U(\vec{r}_i)=\sum_{j \neq i}\left[C_r e^{-|\vec{r}_i-\vec{r}_j|/l_r} - C_a e^{-|\vec{r}_i-\vec{r}_j|/l_a} \right],
\label{Eq_potential}
\end{eqnarray}
where the parameter pairs ($C_a$,$l_a$) and ($C_r$,$l_r$) specify the respective amplitudes and ranges of the attractive and repulsive terms of the Morse potential. The SPPs model presented in Eqs. (\ref{Eq_1_model})-(\ref{Eq_2_model}) embodies the generic components necessary for swarming behavior, namely particle separation, cohesion, and velocity matching \cite{Reynolds1987}. Depending on the interaction parameters, the model exhibits qualitatively different collective motions \cite{Orsogna2006}. We fix $C_a=0.5$, $l_a=2.0$, $C_r=1.0$, and $l_r=0.5$ corresponding to a regime in which the collective motion is consistent with swarming \cite{Orsogna2006}. In addition, the applicability of this model has been proved in numerous studies of swarm's collective behavior \cite{Chuang2007,Strefler2008,Nguyen2012,Armbruster2017,Hindes2021,Hindes2021b,Yang2022}.

The dynamics of swarms of self-propelled particles depend on the interplay between their energy admission and dissipation \cite{Ramaswamy2017}. Here, this aspect is captured by the effective friction function (first term in Eq. (\ref{Eq_2_model})). This function has a zero at $\vec{v}_c^2=\frac{\alpha}{\beta}$, representing an attractive characteristic velocity that defines two velocity ranges: $|\vec{v}_i|<|\vec{v}_c|$, where particles are accelerated, and $|\vec{v}_i|>|\vec{v}_c|$, where they are damped \cite{Schweitzer1998,Ebeling1999,Schweitzer2001}. Once the velocity $|\vec{v}_c|$ is established, the effective friction function vanishes, and the swarm dynamics becomes Hamiltonian \cite{Schweitzer1998,Ebeling1999,Orsogna2006}. However, the particle swarms approach the velocity $\vec{v}_c$ in two distinct self-organized configurations: i) A rotational state (RS), where the particles rotate around a common center without global translational motion. ii) A translational state (TS), where the particles travel with constant velocity in a crystallized formation. Interestingly, the approach to one state or another depends on the swarm's initial conditions (ICs); that is, each state has a basin of attraction with its respective basin boundaries. Additionally, depending on the ICs, both states can display various internal features that we do not differentiate in this study. For instance, in the RS, the swarm can rotate clockwise or counterclockwise, while in the TS, it can move in any direction within physical space. For additional types of different states coexisting in the state space of systems with swarm behavior, see Refs. \cite{Orsogna2006,Chuang2007,Aranson2022, Aguiar2023,Yadav2023exotic}. Once the ICs are specified, the swarm goes through a transient phase of motion during which the effective friction is still active. After reaching one of the states, the particle swarms remain in it with their respective collective state; in other words, RS and TS are absorbing states.

\begin{figure}[!h]
\centering
\includegraphics[width=8.3cm,height=4.4cm]{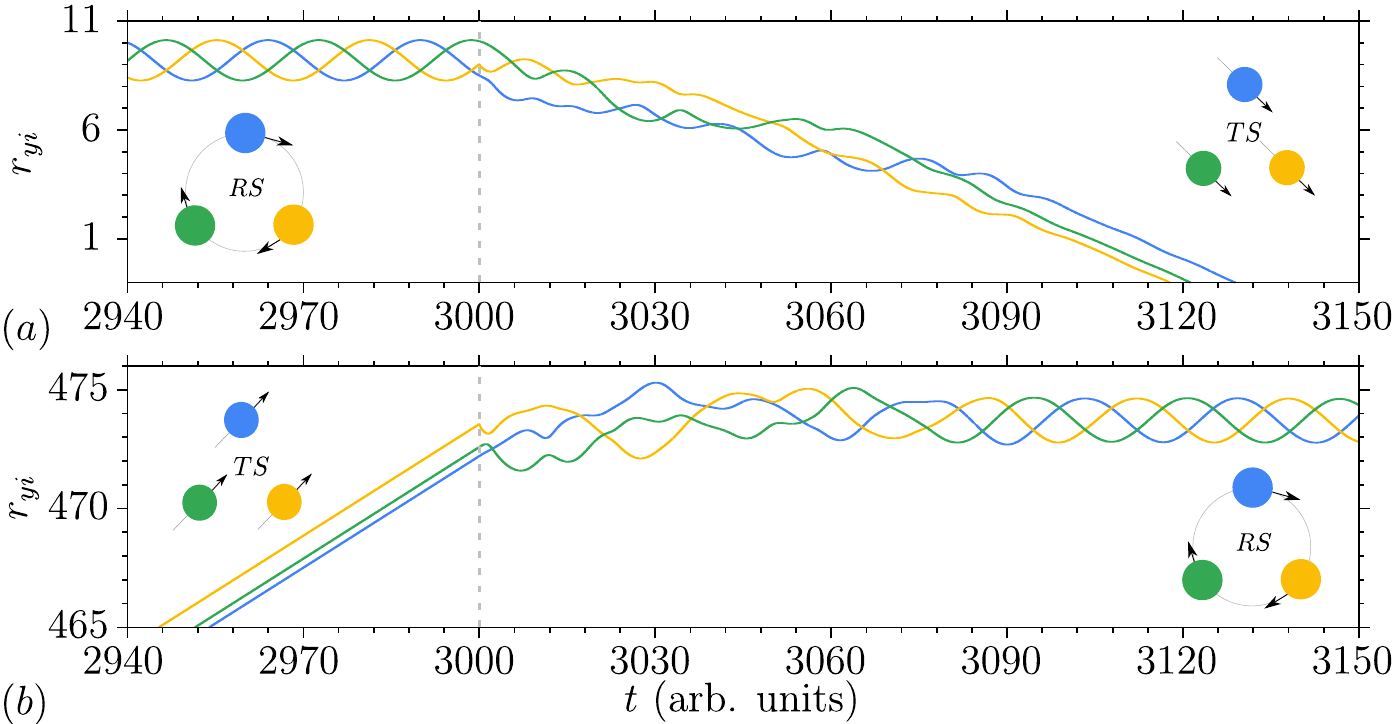}
\caption{Time evolution of the SPPs system in Eqs. (\ref{Eq_1_model})-(\ref{Eq_2_model}) for $N=3$. An acceleration $\vec{a}_i$ is applied to one of the particles at $t=3000$ (arb. units). (a) For $\vec{a}_3=(1.565, 182.43^{\circ})$, a transition is triggered from RS to TS. (b) For $\vec{a}_3=(1.361, 191.27^{\circ})$, a transition from TS to RS occurs. Swarm energy parameters are $\alpha=0.15$ and $\beta=3.0$.}
\label{figure_1}
\end{figure}

We first consider a swarm with $N=3$ particles to capture the essential features of both states. For this case, the particles of swarms in the RS are spatially arranged in a ring rotating with a velocity given by $|\vec{v}_i|=|\vec{v}_c|$. The velocity of the center of mass is $\vec{v}_{cm}=0$, see left-hand side of Fig. \ref{figure_1}(a). Conversely, for swarms in the TS, the particles travel to infinity in a triangle formation with constant individual velocities $|\vec{v}_i|=|\vec{v}_c|$, resulting in $\vec{v}^2_{cm}=\alpha / \beta$ shown in the left-hand side of Fig. \ref{figure_1}(b). Following these considerations on the swarm's possible collective states, we proceed to our control strategy which consists of applying an instantaneous acceleration $\vec{a}_i$ locally to one of the particles for swarms occupying either states. This acceleration is defined as $\vec{a}_i=(\Delta v, \theta^{\circ})$, where $\Delta v$ indicates its magnitude, and $\theta^{\circ}$ represents its orientation in a polar diagram centered at the particle's velocity $\vec{v}_i$ immediately before the acceleration. After this local action, the swarms in the RS (TS) undergo a global transition to the TS (RS), as shown in Figs. \ref{figure_1}(a) and \ref{figure_1}(b), respectively. During the transitions to either collective state, the particles display a transient irregular phase of motion, as seen in the middle portions of Figs. \ref{figure_1}(a) and \ref{figure_1}(b). Such irregular motion arises from the complexities of the boundaries separating the basins of attraction of these states. In the literature, an irregular transient motion has recently been observed in systems of self-propelled elements \cite{Pikovsky2021,Aranson2022}. In addition, transitions between RS and TS have been previously observed only in stochastic systems with random noise applied to all swarm particles \cite{Mikhailov1999,Erdmann2002,Erdmann2005}.

\begin{figure}[!htp]
\includegraphics[width=8.5cm,height=3.15cm]{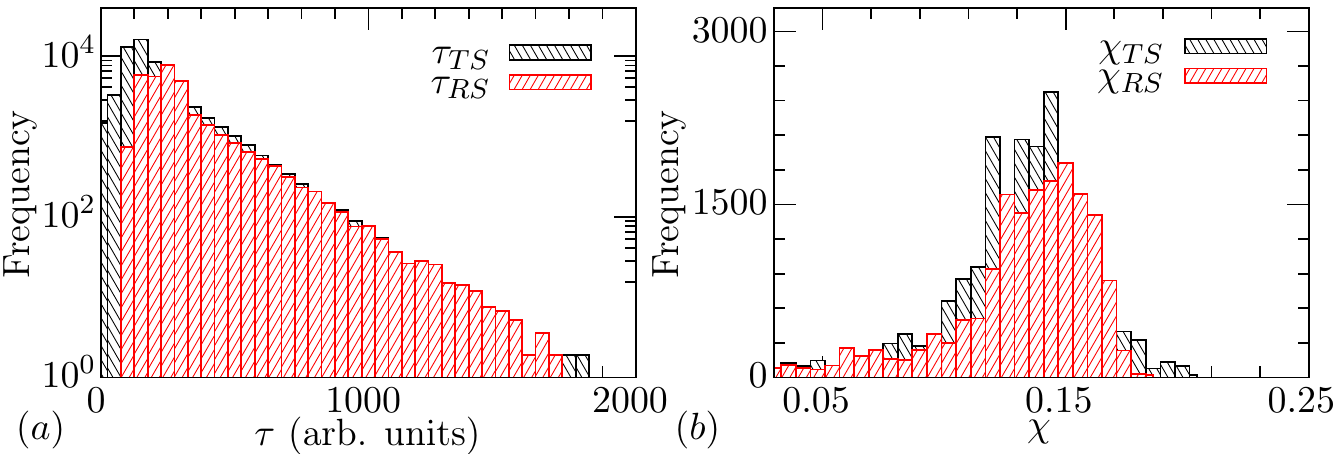}
\caption{For swarms with $N=3$ particles: (a) Distribution of transition times $\tau$ for swarms originally in RS (black) and TS (red). (b) LLE $\chi$ during the transition from RS (black) and from TS (red). Swarm energy parameters are $\alpha=0.15$ and $\beta=3.0$.}
\label{figure_2}
\end{figure}

To gain a deeper understanding of the irregular motion during transitions to alternative collective states, we examine sets of the local accelerations $\vec{a}_i$ that trigger these transitions for statistical analysis. First, we demonstrate that the probability distribution of the transition duration $\tau$ follows an exponential distribution, which indicates random behavior [Fig. \ref{figure_2}(a)]. The mean transition time from RS to TS, and vice versa, is $\langle \tau \rangle \approx 190$ (arb. units). In addition, we estimate the largest Lyapunov exponents (LLE) corresponding to the transition times. We verify that the average LLE is positive for transitions in both directions and approximately the same $\langle \chi_L \rangle \approx 0.15$ [Fig. \ref{figure_2}(b)], indicating transient chaotic behavior \cite{Lai2011}. This chaotic motion points to an unstable chaotic set (chaotic saddle) embedded in the basin boundaries of the collective states and mediates the transition in both directions. Chaotic saddles have been identified at the border of a variety of dynamical behaviors \cite{Skufca2006,Eckhardt2007,Rempel2007,Joglekar2015,Ansmann2016,lilienkamp2017,Lucarini2019,Medeiros2021}. In general the basin boundaries containing embedded chaotic saddles are fractal and are usually extended over large portions of the state space \cite{Hsu1988,Poon1995,Macau1999}. This feature is an important ingredient for the local control strategy proposed here once it allows trajectories of individual particles in a given collective state to easily access the basins of the opposite state via the application of the local accelerations $\vec{a}_i$, causing the entire swarm to switch its collective dynamics.

\begin{figure}[!htp]
\centering
\includegraphics[width=8.5cm,height=2.9cm]{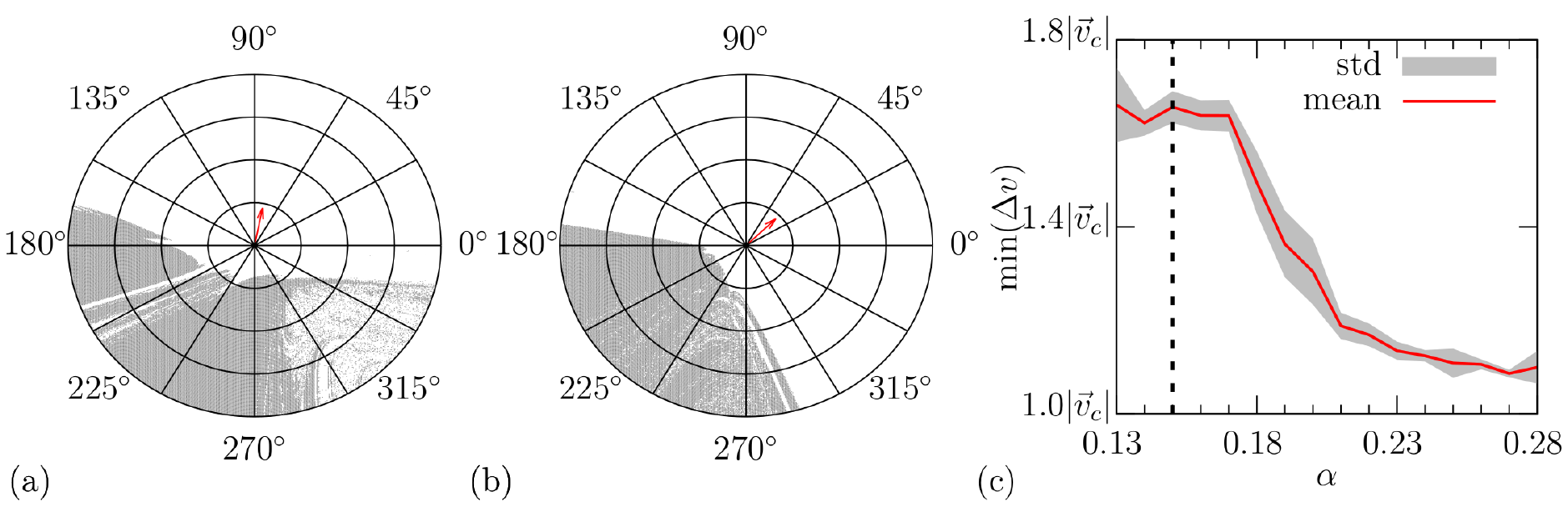}
\caption{For swarms with $N=3$ particles, polar diagrams ($\Delta v$, $\theta$) illustrating the acceleration set $A_3$ applied to particle $i=3$ originally in (a) RS and (b) TS. The set of effective accelerations $A^{eff}_3$ is marked in gray. The red arrow represents the direction of $\vec{v}$ right before the acceleration is applied. Swarms energy parameters are $\alpha=0.15$ and $\beta=3.0$. (c) Minimum amplitude $\Delta v$ of the effective accelerations $\vec{a}_3$ as a function of the energy admission $\alpha$ applied in seven realizations of swarms originally in RS with dissipation fixed at $\beta=3.0$. The dashed line marks $\alpha=0.15$.}
\label{figure_3}
\end{figure}

Next, we further investigate the sets of local accelerations $\vec{a}_i$ that can cause collective transitions. For that, we formally define the set $A_i=\{\vec{a}_i \in \mathbb{R}^2| \Delta v \in [0,2], \theta \in [0,360^{\circ}]  \}$ composed of $M_i$ applied accelerations equally spaced in this interval. We call $A^{eff}_{i}$ the subset of $A_i$ composed of those accelerations that effectively initiate a transition. In Fig. \ref{figure_3}, we show the set $A_i$ of accelerations $\vec{a}_i=(\Delta v, \theta)$ applied to the particle $i=3$ at $t=3000$ (arb. units). The set $A^{eff}_{3}$ is marked in gray. For swarms in the RS, we observe that the set $A^{eff}_{3}$ of accelerations causing the transition into the TS spans an angle $\theta \approx 180^{\circ}$, featuring fractal boundaries and oriented in the opposite direction of the movement of particle $i=3$ (red arrow) [Fig. \ref{figure_3}(a)]. Due to the rotational properties of the particles in the RS, where the red arrow changes its direction over time, the orientation of $A^{eff}_{3}$ in the polar diagram also undergoes rotation, consistently opposing the particle's movement direction. In this example, the minimum amplitude $\Delta v$ of the acceleration $\vec{a}_3$ required for a transition to occur is min$(\Delta v) \approx 1.6|\vec{v}_c|$. For swarms in the TS, we verify that the set of $A^{eff}_3$ causing the transition into RS spans an angle $ \theta \approx 120^{\circ}$ with minimum amplitude $\Delta v \approx 1.6|\vec{v}_c|$ also containing fractal boundaries [Fig. \ref{figure_3}(b)]. The orientation of $A^{eff}_3$ in the polar diagram is also opposite to the movement of the accelerated particle ($i=3$). Since that in TS, the particles preserve movement direction, the orientation of $A^{eff}_{3}$ is also kept the same over time. With this in mind, we emphasize that the correlation between the direction of movement of accelerated particles and the orientation of the sets $A^{eff}_{i}$ offers insights into determining the accelerations $\vec{a}_i$ that are more likely to initiate a transition. Moreover, the fractal boundaries observed in these polar diagrams are consistent with the transient chaotic motion resulting from the chaotic saddle in the system's state space. The features of this chaotic saddle, which mediates the transitions, and the size of the basins of attraction for both states, are influenced by the parameters that regulate the energy balance of the particles. As a result, the minimum amplitude of the effective accelerations also varies with these parameters. This fact is evident in Fig. \ref{figure_3}(c), where the minimum amplitude, min($\Delta v$), is determined as a function of the energy admission parameter $\alpha$ for seven realizations with different ICs of the swarm in the RS while keeping the dissipation fixed at $\beta=3.0$. In this figure, we observe that increasing the energy input into the particles reduces the minimum acceleration amplitude required to induce a transition from RS to TS. Therefore, the energy balance of the particles plays an essential role in the local controllability of the swarms.

\begin{figure}[!htp]
\centering
\includegraphics[width=8.4cm,height=3.0cm]{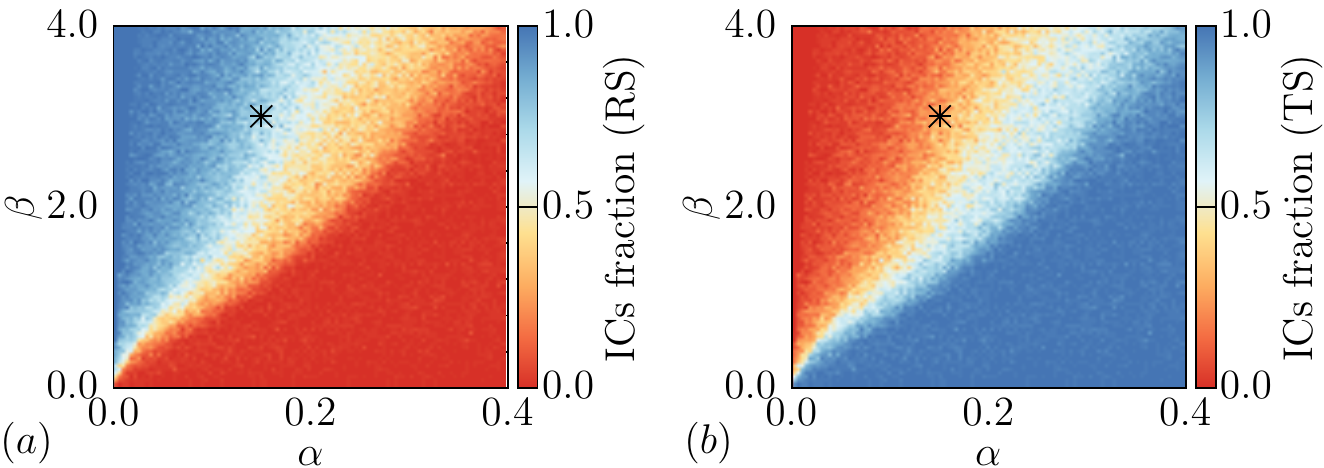}
\caption{Parameters plane ($\alpha$, $\beta$) corresponding to the constants governing the energy admission and dissipation of Eqs.~(\ref{Eq_1_model})-(\ref{Eq_2_model}), respectively. Each parameter pair has $60$ realizations of the swarm with different initial conditions (ICs). The color code corresponds to the normalized fraction approaching (a) RS and (b) TS. The black star symbols mark the swarms energy parameters $\alpha=0.15$ and $\beta=3.0$.}
\label{figure_4}
\end{figure}

Furthermore, adjusting the system parameters to enable the coexistence of different collective states, each corresponding to alternative swarm behaviors, is a prerequisite for the effectivity of the proposed local control $\vec{a}_i$. Considering this, it becomes necessary to identify the swarm parameters that facilitate the coexistence of both the RS and TS states. A detailed analysis of the parameters composing the potential function $U(\vec{r}_i)$, which controls the interactions among the particles, can already be found in Ref. \cite{Orsogna2006}. Therefore, we complement this analysis by obtaining the region of bistability between RS and TS in a parameter space defined by the parameters governing energy admission ($\alpha$) and the one controlling dissipation ($\beta$). In this analysis, the parameters that compose $U(\vec{r}_i)$ are fixed to the previously defined values. Hence, in Fig.~\ref{figure_4}, we consider the parameter intervals $\alpha \in [0.0,0.4]$ and $\beta \in [0.0,4.0]$ and perform $60$ realizations of the swarm with different ICs randomly chosen in the intervals $\vec{r}_{0i} \in [-2.0,2.0]$ and $\vec{v}_{0i} \in [-2.0,2.0]$. The color code indicates the fraction of ICs that converge to the RS [Fig.~\ref{figure_4}(a)] and the TS [Fig.~\ref{figure_4}(b)]. In these figures, the interval with lighter color shades (light blue, white, and light red) corresponds to the parameters where both collective states occur with similar probability. We verify that this interval occupies a significant region of the considered parameter plane, evidence of the structural robustness of the proposed local control strategy. The red cross indicates the parameters $\alpha=0.15$ and $\beta=3.0$ utilized in simulations across this study.

\begin{figure*}[!htp]
\centering
\includegraphics[width=17.4cm,height=6cm]{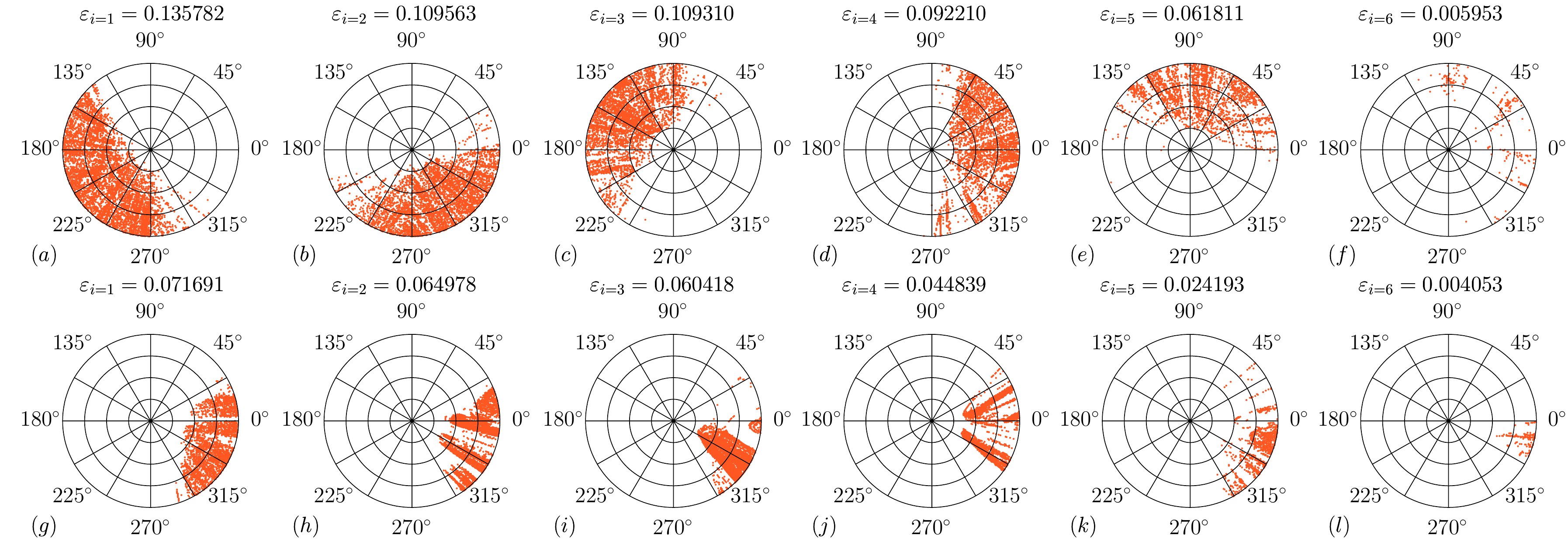}
\caption{For swarms originally in the RS: (a)-(f) Polar diagram ($\Delta v$, $\theta$) of accelerations applied to each particle of independent realizations of swarms with $N=6$. (g)-(l) For swarms originally in the TS. The red dots indicate the set of effective accelerations, $A^{eff}_i$, and $\varepsilon_i$ corresponds to the respective control effectivity of the particle $i$. The accelerations are applied at $t=1500$ (arb. units), and the total number of accelerations per particle is $M_i=7894$. Swarm energy parameters are $\alpha=0.15$ and $\beta=3.0$.}
\label{figure_5}
\end{figure*}

We now investigate the sets of effective accelerations $A^{eff}_i$ in more detail. We first show $A^{eff}_i$ in polar diagrams for accelerations applied at $t=1500$ (arb. units) to each particle of swarms with $N=6$. The sets $A^{eff}_i$ of each particle $i$ ($i=1, \dots, 6$) are marked in red in Figs.~\ref{figure_5}(a)-\ref{figure_5}(f) for swarms initially in the RS and in Figs.~\ref{figure_5}(g)-\ref{figure_5}(l) for swarms initially in the TS. We emphasize that the size of the sets $A^{eff}_i$ varies among the particles in both collective states. To gain further insights into the size of these sets, we quantify their relative volume as $\varepsilon_i = \frac{Vol(A^{eff}_i \cap A_i)}{Vol(A_i)}$ \cite{Menck2013}. We estimate this measure by counting the number of effective accelerations, denoted as $M^{eff}_i$, and dividing it by the total number of accelerations, denoted as $M_{i}$, applied to each particle $i$ in a swarm of size $N$. Since $\varepsilon_i$ represents the fraction of effective acceleration of a given particle, it reflects its ability to induce transitions between different collective states. Therefore, we refer to $\varepsilon_i$ as the control effectivity of particle $i$. In Fig.~\ref{figure_5}, the values of $\varepsilon_i$ corresponding to each particle are indicated on their respective polar diagrams.

\begin{figure}[!htp]
\centering
\includegraphics[width=8.4cm,height=3.0cm]{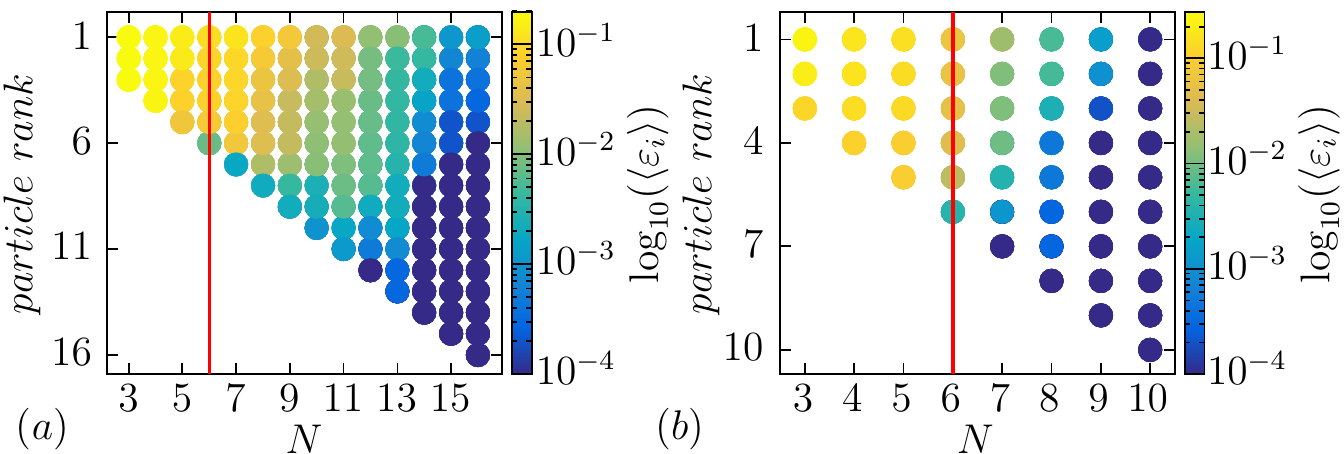}
\caption{Average control effectivity $\langle \varepsilon_i \rangle$ (color code) with $i=1,\dots,N$ over $7$ different sets of ICs as a function of the particle rank and the swarm size $N$. (a) For the RS. (b) For the TS. The red line corresponds to case studied in Fig.~\ref{figure_5}. The accelerations are applied at $t=1500$ (arb. units), and the total number of accelerations per particle is $M_i=7894$. Swarms energy parameters are $\alpha=0.15$ and $\beta=3.0$.}
\label{figure_6}
\end{figure}

The values of $\varepsilon_i$ displayed in Fig.~\ref{figure_5} for both swarm's collective states exhibit significant variations among the particles, suggesting the existence of an internal hierarchy of control effectivity. To confirm this possibility, we calculate $\varepsilon_i$ for particles in various swarm sizes $N$ and multiple swarm realizations with randomly selected initial conditions (ICs). For a given $N$, we rank the particles that make up the swarm's collective state in descending order of their $\varepsilon_i$ values. Subsequently, we calculate the average of the ordered values of $\varepsilon_i$ over seven different sets of ICs to obtain $\langle \varepsilon_i \rangle$. In Fig.~\ref{figure_6}, we present $\langle \varepsilon_i \rangle$ (color-coded) as a function of particle rank and swarm size. We confirm that $\langle \varepsilon_i \rangle$ decreases as the swarm size increases, reaching the minimum fraction detectable by our estimation, $1/M_i$, at $N=16$ for swarms in the RS [Fig.~\ref{figure_6}(a)] and $N=10$ for swarms in the TS [Fig.~\ref{figure_6}(b)]. The faster decrease in $\langle \varepsilon_i \rangle$ indicates that the TS is less vulnerable to local accelerations. Furthermore, for both collective states, we notice that $\langle \varepsilon_i \rangle$ consistently decreases with respect to particle rank, illustrating the hierarchical organization of particles within the swarms.

\begin{figure}[!htp]
\centering
\includegraphics[width=8cm,height=9.5cm]{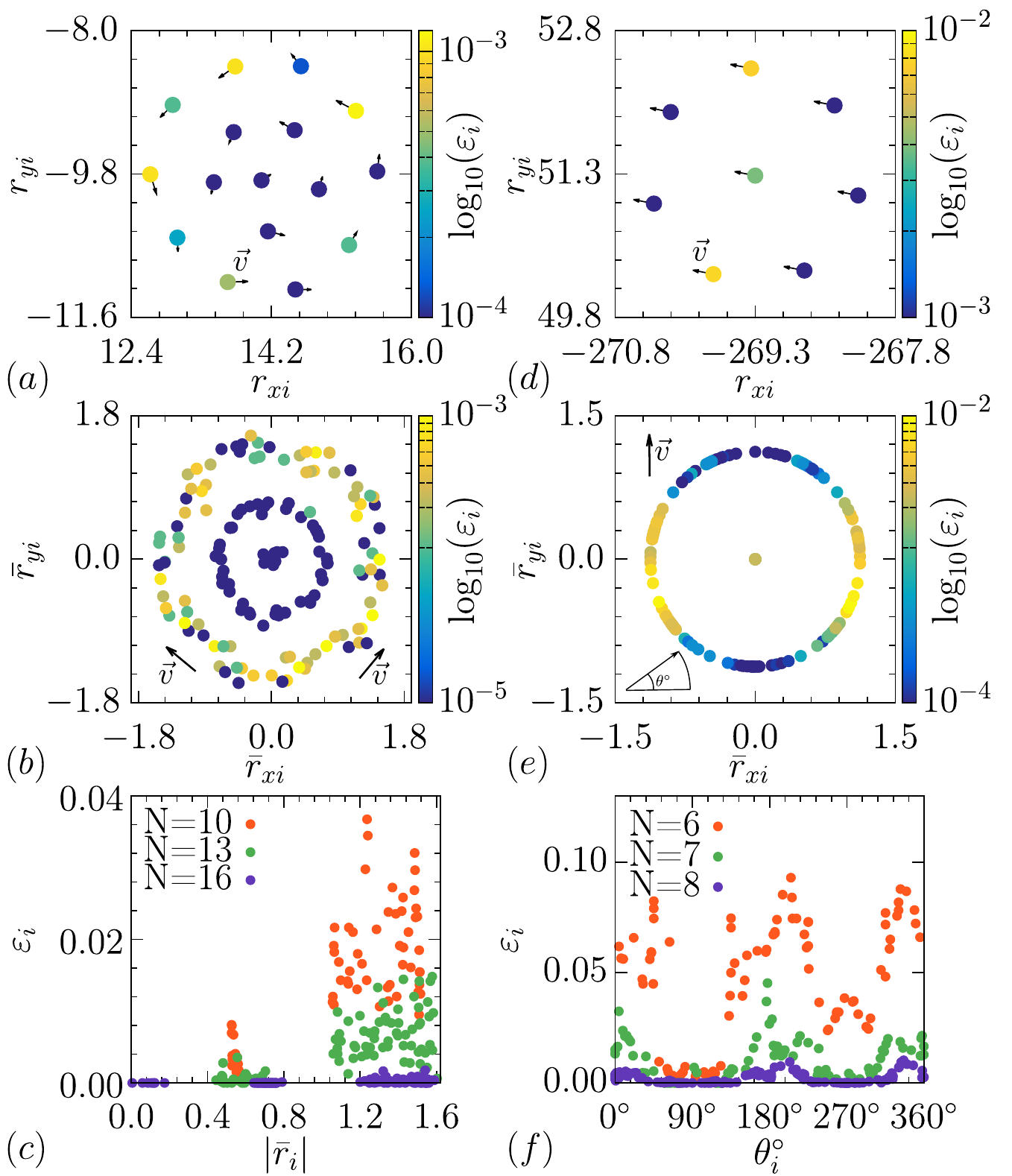}
\caption{(a) Positions of the particles in a swarm with $N=16$ particles in the RS. The color code indicates each particle's control effectivity, $\varepsilon_i$, while the arrows represent the particle's velocity, $\vec{v}_i$. (b) Seven different realizations of swarms in the RS, each with different ICs. (c) Control effectivity of the particles, $\varepsilon_i$, as a function of the distance to the center of mass, $|\bar{r}_i|$, of the RS for various swarm sizes, $N$. (d) Positions of the particles in a swarm with $N=8$ particles in the TS. (e) Seven different realizations of swarms in the TS, each with different ICs. (f) $\varepsilon_i$ as a function of the angle, $\theta^\circ$ (as shown in (e)), for varying swarm sizes, $N$. The accelerations are applied at $t=1500$ (arb. units), and the total number of accelerations per particle is $M_i=7894$. Swarms energy parameters are $\alpha=1.0$ and $\beta=8.0$. Swarms energy parameters are $\alpha=0.15$ and $\beta=3.0$.}
\label{figure_7}
\end{figure}

Next, we untangle these hierarchies by examining the positions of the particles with a specific $\varepsilon_i$ within the swarms at the instant right before the local acceleration is applied. In Fig. \ref{figure_7}(a), in the case of a swarm consisting of $N=16$ particles in the RS, we observe that the outer layer of the formation concentrates the particles with the highest values of $\varepsilon_i$ (color-coded). By contrast, in the case of a swarm consisting of $N=8$ particles in the TS, we observe that particles with the highest $\varepsilon_i$ occupy the axis of the swarm perpendicular to the direction of movement [Fig. \ref{figure_7}(d)]. To ascertain whether these observations represent statistically significant patterns, it is crucial to investigate swarms of different sizes and initial conditions (ICs). In such a statistical approach, the swarms exhibit various qualitative features, including different directions of rotation for the RS and direction of translation for the TS. To ensure comparability across realizations, we track the particles by their positions ($\bar{r}_{xi}$, $\bar{r}_{yi}$) relative to the swarm's center of mass \cite{note2}. Using these coordinates, we confirm the emergence of a distinct hierarchical pattern with particles possessing high $\varepsilon_i$ occupying the outer layers of the RS in seven different sets of initial conditions [Fig. \ref{figure_7}(b)]. To highlight this pattern across various swarm sizes, we illustrate that $\varepsilon_i$ increases with larger distances from the swarm's center of mass, $|\bar{r}_i|$, for $N=10$, $N=13$, and $N=16$ [Fig. \ref{figure_7}(c)]. Now, in the case of swarms in the TS, we verify that particles with the highest $\varepsilon_i$ are distributed along the axis oriented perpendicular to the direction of motion in seven different sets of ICs [Fig. \ref{figure_7}(e)] \cite{note1}. For different $N$, we observe that $\varepsilon_i$ peaks around the angles $\theta^{\circ}=0^{\circ}$ and $\theta^{\circ}=180^{\circ}$ (measured as shown in Fig.~\ref{figure_7}(e)) for $N=6$, $N=7$, and $N=8$ [Fig. \ref{figure_7}(f)].

In order to gain insights into the emergence of hierarchies of control effectivity among the particles and patterns of high control effectivity within the swarms observed in Fig.~\ref{figure_6} and Fig.~\ref{figure_7}, we must consider that the success of the local control strategy proposed here depends on the global stability of the collective states. Specifically, the effectivity of instantaneously accelerating one particle to switch the swarm's collective state relies on the ability of this acceleration to force swarm trajectories to traverse the basins of attraction of the original collective state and reach the basin of attraction of the alternative state. However, the required amplitude of these local accelerations for switching typically depends on specific features within the cross-section of the system's basins of attraction in the direction of the perturbed particle. Interestingly, one of the most significant features within these cross-sections that can facilitate the action of our control is the distance between the collective state itself and the boundaries of its basin of attraction. In the case of the SPPs model studied here, where the basin boundaries form a complex fractal set, the distance to the collective states varies across the cross-sections corresponding to different particles. This cross-section variability gives rise to the hierarchy of control effectivity among the particles observed here. In turn, we attribute the formation of patterns of high control effectivity to the eventual concentration of particles for which the corresponding basin cross-section encompasses a shorter distance between the state and the basin boundaries. This concentration of particles in a specific region of the swarm's spatial conformation may arise from macroscopic features of the particular swarm region, such as the velocity of the particles for the RS or the number of neighbors for the TS. While it can be challenging to generalize patterns of high controllability, their understanding, in conjunction with the concept of particle hierarchies, offers fresh perspectives for controlling artificial swarms.

\begin{figure}[!htp]
\centering
\includegraphics[width=8.4cm,height=3.1cm]{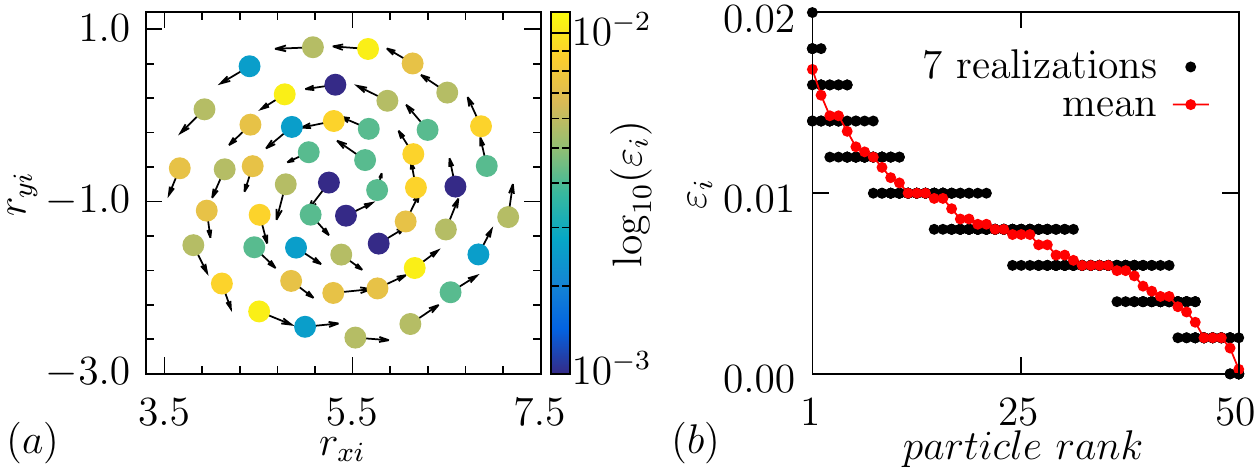}
\caption{(a) The positions of the particles in a swarm with $N=50$ particles in the RS are shown. The color code indicates each particle's control effectivity $\varepsilon_i$, while the arrows represent the particle's velocity $\vec{v}_i$ with $i=1,\dots,N$. (b) Control effectivity $\varepsilon_i$ is plotted as a function of the particle's rank within seven realizations of the swarm with $N=50$ particles, each starting with different initial conditions (ICs). The accelerations are applied at $t=1500$ (arb. units), and the total number of accelerations per particle is $M_i=500$. Swarms energy parameters are $\alpha=1.0$ and $\beta=8.0$.}
\label{figure_8}
\end{figure}

Now, we would like to point out that by appropriately tuning the parameters $\alpha$ and $\beta$, which define the energy input and dissipation for the particles, it is possible to locally control even larger swarms than the ones shown in Fig.~\ref{figure_6}. To illustrate this finding, we fix $\alpha=1.0$ and $\beta=8.0$ to obtain the control effectivity $\varepsilon_i$ for particles in a swarm of size $N=50$ in RS. Hence, in Fig.~\ref{figure_8}(a), we show the particle positions $(r_{xi}, r_{yi})$ for one realization of the RS, indicating the control effectivity $\varepsilon_i$ of the local accelerations $\vec{a}_i$ with a color code. We can not identify a distinct spatial pattern of control effectivity among the particles in this parameter set. Nevertheless, in Fig.~\ref{figure_8}(b), where we present results from seven realizations with different ICs, we continue to observe the presence of internal hierarchies among the particles, as previously demonstrated in smaller swarms.

\begin{figure}[!htp]
\centering
\includegraphics[width=8.4cm,height=3.1cm]{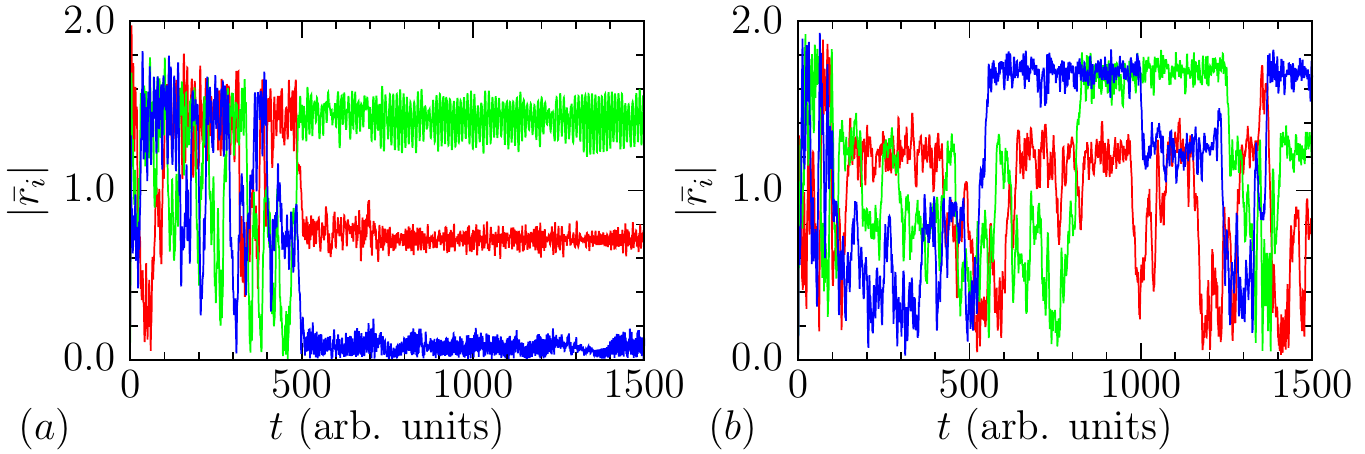}
\caption{Time evolution of the distance to the swarm's center of mass $|\bar{r}_i |$ of three particles marked in different colors for a swarm of size: (a) $N=16$ particles with energy parameters $\alpha=0.15$ and $\beta=3.0$. (b) $N=50$ particles with energy parameters $\alpha=1.0$ and $\beta=8.0$.}
\label{figure_9}
\end{figure}

We attribute the absence of a spatial pattern of control effectivity in the swarm with $N=50$ particles to the irregular movement of the particles across the swarm formation at the instant $t=1500$ (arb. units) in which the control is applied. More specifically, the particles are not organized in well-defined layers for the larger swarm, as seen in the cases presented in Fig.~\ref{figure_7}. To elucidate this point, in Fig.~\ref{figure_9}, we show the time evolution of the distance to the swarm's center of mass $|\bar{r}_i|$ of three different particles belonging to the swarm with $N=16$ particles [Fig.~\ref{figure_9}(a)] and with $N=50$ particles [Fig.~\ref{figure_9}(b)]. Notice that for $N=16$ particles, the time evolution of $|\bar{r}_i|$ reaches a plateau corresponding to the rotating motion of each observed particle in a different, well-defined layer. Conversely, for $N=50$ particles, the measure $|\bar{r}_i|$ varies significantly in time, indicating that the three observed particles transit across the spatial extension of the swarm, not staying confined in any well-defined rotating layer. Therefore, local accelerations can still control large swarms, but the possible existence of invariant patterns of high control effectivity requires further investigation of their internal dynamics.

\section{Homoclinic intersections in a reduced SPP model}

\begin{figure}[!htp]
\centering
\includegraphics[width=8.5cm,height=3.28cm]{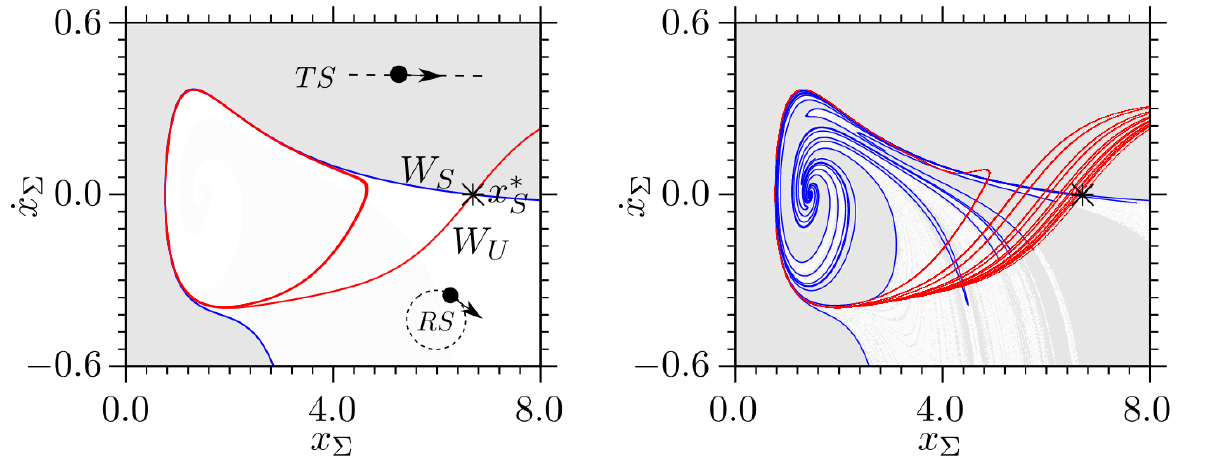}
\caption{Basins of attraction at the cross-section $\Sigma$. The white color corresponds to RS, while gray to TS. The blue (red) curve is the stable (unstable) manifold of the saddle point $x^{*}_S$. (a) Smooth basin boundaries for $\sigma=0.0001$. (b) Fractal basin boundaries for $\sigma=0.0005$. Parameters: $\alpha=0.2$ and $\beta=2.0$.}
\label{figure_10}
\end{figure}

We now show that homoclinic intersections of the stable and unstable manifolds \cite{MCdonald1985,Moon1985} are a possible mechanism giving rise to such fractal boundaries, which play an important role in the control effectivity. We consider a low-dimensional SPPs system by perturbing the motion of one particle in a periodically perturbed potential to represent its coupling with other particles. The resulting equation of motion is:
\begin{equation}
 \ddot{x}=(\alpha - \beta \dot{x}^2)\dot{x} - \frac{\partial U'(x,t)}{\partial x},
 \label{eq_low_dimensional}
\end{equation}
where the interaction is given by $U'(x)=U(x) + xP(t)$ consisting of an one-dimensional Morse potential $U(x)$ (Eq.~(\ref{Eq_potential})) and a periodic forcing $P(t)=\epsilon+\sigma sin(\omega t)$. The frequency $\omega=0.2834$ mimics the oscillations of the other particles, corresponding to the frequency of the particle`s oscillations in an unperturbed potential. The parameters $\epsilon=0.0088$ and $\sigma$ control the perturbation's overall amplitude and the periodic forcing's amplitude, respectively. Hence, Eq.~(\ref{eq_low_dimensional}) yields a three-dimensional state space $\Re^2 \times \mathcal{S}^1$. Since $\mathcal{S}^1$ is a periodic component, a stroboscopic mapping is obtained by fixing a cross-section $\Sigma$ at ($x$,$\dot{x}$,$0$). In Fig.~\ref{figure_10}, we show the basins of attraction at $\Sigma$. White corresponds to ICs converging to the RS, while gray represents the ICs converging to the TS. The boundary between these two basins of attraction is given by the stable manifold $W_S$ of the saddle point $x^{*}_S$; (blue curve). The red curve represents the unstable manifold $W_U$ of $x^{*}_S$. For $\sigma=0.0001$, [Fig.~\ref{figure_10}(a)], the basin boundary is smooth. The manifolds $W_S$ and $W_U$ are close but do not touch each other. For $\sigma=0.0005$, [Fig.~\ref{figure_10}(b)], the manifolds have intersected at infinitely many points as homoclinic intersections, creating the fractal basin boundaries. Notice that the fractal basin boundaries occupy a much larger portion of the state space compared to Fig.~\ref{figure_10}(a). This feature enhances the local control effectivity proposed in this study by increasing the likelihood of accelerations reaching the stable manifold of the saddle point. The manifolds were obtained as in \cite{Ciro2018}.

\section{Conclusions}

In summary, we have demonstrated that swarms of interacting self-propelled particles can be locally controlled by applying external accelerations to individual swarm particles. By quantifying the effectivity of accelerating different swarm particles, we have identified hierarchies of controllability in which specific particles have the highest probability of shifting the swarm from a rotational to a translational state and vice versa. For low-density swarms, we have observed that particle hierarchies form patterns of control effectivity within the spatial conformation of the two swarm states. Specifically, for rotational states, we have found that controllability increases with the distance of particles from the center of rotation. In the case of the translational state, particles along the axis transversal to the motion hold a higher position in the hierarchy, influencing control effectivity. These patterns have been observed across various swarm sizes, up to a certain threshold, and under different initial conditions. For denser swarms, we have considered one example of a swarm in the rotational state to demonstrate that local controllability can also be achieved by adjusting the parameters governing the energy input and dissipation of the particles. Although particle hierarchies are also present in this scenario, the patterns of control effectivity are not as clearly discernible. We attribute this behavior to the less ordered movement of particles within the larger swarm.

Furthermore, we have observed chaotic transients during collective state transitions, revealing the existence of fractal sets that separate the swarm's rotational and translational states. We attribute the increase in the likelihood of transitions between these states to this fractal boundary, which extends its coverage across the system's state space. Therefore, these fractal sets play a crucial role in the success of our local acceleration-based control approach for the swarm. In a simplified version of the considered model of interacting self-propelled particles, we demonstrate that the development of fractal basin boundaries occurs through homoclinic intersections, a typical mechanism for boundary fractalization.

The local controllability and the particle hierarchies reported here offer fresh insight into controlling artificial swarms of self-propelled particles in a decentralized manner, in other words, steering the swarm by controlling just one of the self-propelled elements. These findings inspire further research to experimentally verify these hierarchies, evaluate their scalability in higher-density swarms, and assess their robustness to random noise. Additionally, it is worth exploring whether similar hierarchical patterns manifest within collective states in a broader range of systems beyond swarms in physical space.

\begin{acknowledgments}
E.S.M and U.F. acknowledge the support by the Deutsche Forschungsgemeinschaft (DFG) via the project number 454054251 (FE 359/22-1). The simulations were performed at the HPC Cluster CARL, located at the University of Oldenburg (Germany) and funded by the DFG through its Major Research Instrumentation Program (INST 184/157-1 FUGG) and the Ministry of Science and Culture (MWK) of the Lower Saxony State, Germany.
\end{acknowledgments}


\begin{thebibliography}{59}
\bibitem [{not({\natexlab{a}})}]{note2}%
  \BibitemOpen
  \bibfield  {title} {\bibinfo {title} {The coordinates of the center of mass
  are given by \begin{math}r_{xcm}=1/N\sum_i^Nr_{xi}\end{math} and
  \begin{math}r_{ycm}=1/N\sum_i^Nr_{yi}\end{math}. Hence, we
  define the coordinates of the particles with respect to the center of mass as
  \begin{math}\bar{r}_{xi}=r_{xcm}-r_{xi}\end{math}
  and
  \begin{math}\bar{r}_{yi}=r_{ycm}-r_{yi}\end{math}.}\
  }%
\bibitem [{not({\natexlab{b}})}]{note1}%
  \BibitemOpen
  \bibfield  {title} {\bibinfo {title} {For various ICs in the TS, the system moves in different directions within physical space. To ensure comparability of realizations without affecting the interpretation of results, we rotate the spatial components of the center of mass, causing all realizations to move in the same direction in the diagram depicted in Fig. 7(e).}\ }%
\end{thebibliography}
%
\end{document}